# SIZING OF A PV/BATTERY SYSTEM THROUGH STOCHASTIC CONTROL AND PLANT AGGREGATION

Thomas Carriere[1,2], Christophe Vernay[3], Sébastien Pitaval[2,3], François-Pascal Neirac[1], George Kariniotakis[1]
[1]MINES ParisTech, PSL University, Centre PERSEE
CS 10207, rue Claude Daunesse, 06904 Sophia Antipolis Cedex, France
[2]Third Step Energy, 55 allée Pierre Ziller, 06560 Sophia Antipolis Cedex, France
[3]SOLAÏS, 55 allée Pierre Ziller, 06560 Sophia Antipolis Cedex, France

ABSTRACT: The objective of this work is to reduce the storage dimensions required to operate a coupled photovoltaic (PV) and Battery Energy Storage System (BESS) in an electricity market, while keeping the same level of performance. Performance is measured either with the amount of errors between the energy sold on the market and the actual generation of the PV/BESS i.e. the imbalance, or directly with the revenue generated on the electricity market from the PV/BESS operation. Two solutions are proposed and tested to reduce the BESS size requirement. The first solution is to participate in electricity markets with an aggregation of several plants instead of a single plant, which effectively reduces the uncertainty of the PV power generation. The second is to participate in an intra-day market to reduce the BESS usage. To evaluate the effects of these two solutions on the BESS size requirement, we simulate the control of the PV/BESS system in an electricity market.

## 1 INTRODUCTION

There is a significant uncertainty in the future funding of ground-mounted PV power plants. Traditional feed-in tariffs are disappearing, and most support schemes are now based on participation in electricity markets, where PV power producers are financially responsible for their forecast errors. Thus, the actual revenue generated from PV power plants is subject to both electricity price and PV power generation uncertainty. In order to mitigate these uncertainties, Battery Electricity Storage Systems (BESS) are often studied. However, they are still very costly. In this paper, we investigate how participating in intra-day electricity markets, and using an aggregation of PV power plants instead of a single plant can reduce the BESS size requirements. To do so, we simulate the control of a coupled PV/BESS system in the French electricity market environment.

The most used method of control for coupled PV/BESS systems is the Model Predictive Control (MPC). It consists in optimizing the control of the BESS on a receding horizon, so that the forecast future state of the system is taken into account when deriving the command for the immediate future. In most cases, the loss function used in the MPC controller is either the producer's profit [1], [2] or the energy imbalance [3], which is the deviation between the energy sold in the electricity market and the actual PV power generation. Some authors also propose an MPC approach to bid on intra-day market sessions [4], [5], however the intra-day market that they consider is organized in sessions, which is different from the continuous intra-day market that we consider in this paper. The uncertainty of the upcoming PV production is sometimes included in both the day-ahead planning and the real-time control of the BESS, as in [6], [7] or [8]. We also include a modelization of the BESS ageing in our optimization problems as in [9]. While few papers consider the uncertainty of renewable energy forecasts, [10] and [11] model this uncertainty using production scenarios. An MPC controller that included all these elements was proposed in [12] but it did not consider an intra-day market.

The main innovation of this work is the implementation of a general stochastic MPC controller that optimizes the real-time control as well as the day-ahead bidding and the participation in a continuous intra-day market of a coupled PV/BESS system taking into account the uncertainty in the upcoming PV power generation, based on an arbitrary objective function, namely the imbalance or the penalties. Another innovation is the quantification of the added value from using a BESS both in terms of imbalance reduction and revenue maximization.

## 2 MARKET STRUCTURE

In this paper, we study the participation of a PV power plant coupled with a BESS on a day-ahead market and dual-pricing balancing market, with a possibility to sell or buy energy on an intra-day market. The market structure assumed is:

-A day-ahead market where each participant has to submit buying or selling orders the day before delivery.

-An intra-day market where energy can be sold or bought up to 30 minutes before delivery

-A balancing market where each BRP has to take responsibility for its imbalances after delivery.

### 2.1 Day-ahead market

On day-ahead markets, each participant must submit buying or selling bids before the Gate Closure Time (GCT). Then, all the buying and selling bids are combined to derive aggregated demand and supply curves for each PTU of the following day. The intersection of these curves defines the spot market price. The calculation of the spot price after the GCT is called the market clearing.

After the spot price has been calculated, the market participants are nominated for injection on the grid, depending on whether their bid was accepted or not. All selling bids with a price lower than the spot price are fully accepted, and all buying bids with a price higher than the spot price are fully accepted. Buying and selling offers with a price equal to the spot price are partially accepted. Since we are interested in the selling of energy, we will always adopt the point of view of an energy producer in the remaining of the thesis.

In any case, all accepted transactions are settled with the spot price, independently of the initial bid. For example, if a market participant accepts to sell up to 1 MWh for 20 €/MWh and then the spot price is 40 €/MWh after clearing, the participant's bid is fully accepted and he gets 1 MWh x 40 €/MWh = 40 €

### 2.2 Intra-day market

Intra-day markets are physical markets that allow trading electricity after the GCT. They are especially useful for intermittent energy sources that can use updated forecast to correct the positions they took on the day-ahead market.

As for day-ahead markets, they are characterized by a PTU, and a closure time. For example, on the French day-ahead market EPEX Spot, the PTU is the same as the day-ahead market i.e. one hour, and the closure is five minutes before the start of the delivery period.

The pricing can be the same as for day-ahead markets, with an auction mechanism and a settlement price that applies for all participants. However, it is also very common to have pay-as-bid markets, where buying and selling bids are matched as they appear, directly using the bids price. In this paper, the intra-day market is a pay-as-bid market.

In this paper, we assumed that the producer always submits intra-day offers at the spot price, which is known at that time. We also assume that the intra-day offers are always accepted. This is actually false, as the probability of acceptation of the intra-day offers is very dependent on the price of the offer However, we do not have intra-day data that could allow us to simulate the acceptation of intra-day offers. Assuming that intra-day offers are always accepted allows us to estimate the best-case scenario, giving us an upper bound of the value of intra-day markets.

2.3 Balancing market

The balancing market penalizes the imbalance of each producer after delivery. The penalties are proportional to the difference between the energy sold by the participant and the actual energy he injected in the grid, as measured by the TSO. For an electricity buyer, the penalty is the difference between the bought energy and the actual consumed energy. The proportionality coefficient between the imbalance and the penalty is a price derived by the TSO called the balancing price. As a general rule, the penalties $P$ can write:

$$P = \pi_B (E - E_c)$$

Where $\pi_B$ is the balancing price, $E$ is the actual energy injected into the grid and $E_c$ is the energy contracted in the day-ahead and intra-day electricity markets.

In this paper the balancing market is a dual-price balancing market. With dual-pricing rules, there are actually two balancing prices: one for positive imbalances $\pi_+$ and one for negative imbalances $\pi_-$. The balancing price is usually higher than the spot price for negative imbalances and lower than the spot price for positive imbalances, in which case reducing the imbalance is economically beneficial. However, this is not always the case.

2.4 Formulation of the revenue

When making a transaction on the intra-day market, the revenue from the transaction adds to the amount initially sold, and the volume bought or sold adds to the actual production for calculating the imbalance penalty. The complete revenue of a producer that sells an energy $E_c$ on the day-ahead market, then makes a transaction on the intra-day market of an energy volume $E_{ID}$ (positive when energy is bought, negative when energy is sold) for a price $\pi_{ID}$ is:

$$R = \pi_s E_c - \pi_{ID} E_{ID} + (E + E_{ID} - E_c) \pi_B$$

Where $\pi_s$ is the spot price that is given by the market clearing after the bids from all market participants have been submitted.

When considering a BESS, we reformulate by differentiating the part of the production $E$ that comes from the PV panels $E_{PV}$ and the part that comes from the BESS $E_{BESS}$. We also introduce a term $C(E_{BESS})$, that reflects the costs due to aging of the BESS when used to deliver the amount of energy $E_{BESS}$. This is obtained with the rainflow counting algorithm [13]. The aging of the BESS can be divided into two components, i.e. cycling aging and calendar aging, which is the degradation caused by time. In the remainder of the thesis, we will focus on the cycling aging of the BESS and consider its calendar aging as a given lifetime. The end-of-life of the BESS is thus defined as the minimum lifetime given by the cycling and calendar aging. As an example, if the calendar aging gives a lifetime of 20 years, and the cycling aging a lifetime of 50 years, we consider that the actual lifetime of the BESS is 20 years (as opposed to considering that the cycling aging adds up to the 20 years given as the calendar lifetime).

We penalize the revenue with the cost associated with the life-loss of the BESS. Note that the penalized revenue $R'$ is not an actual cash flow, and that the cost associated with the life-loss is only here to make the control of the BESS more conservative regarding the lifetime. The penalized revenue $R'$ then writes:

$$R' = \pi_s (E_{PV} + E_{BESS}) \\ -(E_{PV} + E_{BESS} + E_{ID} - E_c)(\pi_s - \pi_B) \\ - C(E_{BESS})$$

3 PV/BESS CONTROL

3.1 General Control Method

From the market structure, we can see that for each time step of the simulation, we have to take up to three decisions:

-If it is 12 AM, the energy $E_c$ to bid on the day-ahead market for the next day.

-The energy $E_{ID}$ to bid on the intra-day market for the next time step open for the intra-day market.

-The energy $E_{BESS}$ to charge or discharge from the BESS for the next time step.

The method we use to control the PV/BESS system is a Model Predictive Control (MPC). This means that for each time step of the simulation, we update the PV power and price forecasts, and then solve the optimization problem corresponding to each decision on a window including the near future i.e. 12 hours in this paper. Then, we use the first element of the solution for the control of the PV/BESS for the immediate future and move forward to the next time step of the simulation.

More precisely, for each time step of the simulation and for each decision making process, we solve an optimization problem over a time window of $N_{MPC}$ timesteps using the most updated price and PV power forecasts $\hat{E}_{PV}, \hat{\pi}_s, \hat{\pi}_B$:

$$\Theta^* = argmin_{\{\Theta \in R^{N_{MPC}}\}} \sum_{i=1}^{N_{MPC}} L(\Theta_i, \hat{E}_{PV}, \hat{\pi}_s, \hat{\pi}_B)$$

The loss functions $L$ are dependent on the overall objective of the simulation, and are detailed in the next subsections.

All the optimization problems are performed in a stochastic manner relative to the uncertainty in PV power generation. In other words, instead of directly minimizing the loss function, several scenario of PV power production are generated using the method from [14], and then the empirical expected value of the loss function derived from the scenario is minimized. To compare stochastic and deterministic optimizations, we use only the expected value of the PV power generation as a single scenario to perform the deterministic optimization.

With this method, the near future is taken into account when controlling the PV/BESS. This is especially useful when controlling the BESS. For example, if a large amount of energy was sold on the day-ahead electricity market in the evening, using the BESS to compensate forecast errors during the day could lead to prematurely emptying the BESS and thus not being able to fulfill the day-ahead planning. Considering the near future with a MPC controller prevents this situation.

Since the three decision processes (day-ahead offering, intra-day offering and real-time control) are consecutive, each optimization can use the optimal solution from the previous processes.

3.2 Imbalance minimization

When the overall objective of the simulation is to minimize imbalances, the loss functions for each decision process if the imbalance e.g. the difference between the amount of energy sold and the actual energy production.

3.2.1 Day-ahead offering strategy

In that case, the decision must be taken a 12 AM for all the 24 hours of the next day. Thus, we must solve:

$$E_c^* = argmin_{\{E_c \in R^{24}\}} \sum_{i=1}^{24} |E_i - E_{c,i}|$$

Where $E$ is the total output of the PV/BESS. Assuming that we have a forecast $F$ of the Cumulative Distribution Function (CDF) of the power generation $E_{PV}$, we can prove that the optimal solution is given by:

$$E_c^* = F^{-1}\left(\frac{1}{2}\right)$$

A trivial solution is then to not use the BESS in the day-ahead phase, so that the forecast CDF $F$ is the actually the CDF of the PV power generation, and then set the bids following the above equation.

3.2.2 Intra-day offering strategy

To minimize the imbalances, the optimal solution on the intra-day market is to cancel the imbalance using our best expectation of the PV power production. Thus, the intra-day offers are given by:

$$E_{ID}^* = E_c^* - F^{-1}\left(\frac{1}{2}\right)$$

3.2.3 Real-time control

To minimize the imbalances, we use the absolute imbalance as the loss function $L$. Thus we must solve:

$$E_{BESS}^* = argmin_{\{E_{BESS} \in R^{N_{MPC}}\}} \left[\sum_{i=1}^{N_{MPC}} |E_{PV} + E_{ID}^* + E_{BESS} - E_c^*|\right]$$

Since we use the BESS at this stage, we must also consider the operational constraints of the BESS. To define the constraints, we note as $SOC$ (for State of Charge) the amount of energy in the battery at a given time step, relative to its capacity $Cap$.

$$\frac{1}{\eta_{Ch}} Cap(1 - SOC_i) < E_{c,BESS,i} < \eta_{Dis}\, Cap\, SOC_i$$
$$-E_{c,BESS} < \eta_{Ch} E_{PV}$$
$$Cap\, |SOC_i - SOC_{i-1}| \leq K$$

The first constraint ensures that the energy in the BESS is never lower than 0 or higher than the capacity of the battery, taking into account the charge and discharge rates of the BESS, respectively $\eta_{Ch}$ and $\eta_{Dis}$. The second constraint ensures that the BESS can only be charged from the PV plant, and not from the grid. Finally, the third constraint is a limitation on the power rating of the BESS, defined by the parameter $K$.

3.3 Revenue maximization

When the overall objective of the simulation is the actual revenue, the loss function of each optimization problem is the revenue generated by the decision instead of the imbalance. Then, the optimization problems that we have to solve at each step are the following.

3.3.1 Day-ahead offering strategy

When the BESS is used at both the day-ahead and real-time levels, then the entire formulation of the penalized revenue $R'$ is optimized. Once again, we separate the bids into one part accompanied by uncertainty from the PV plant $E_{c,PV}$, and the output from the battery $E_{c,BESS}$. Since the BESS is controllable, we assume that the actual output of the BESS $E_{BESS}$ is always equal to the amount bid $E_{c,BESS}$. With these assumptions, the optimization problem that needs to be solved to derive the optimal bids is:

$$E_{c,PV}^*, E_{BESS}^* = argmax_{\{E_{c,PV} \in R^{24}, E_{c,BESS} \in R^{24}\}} R'(E_{c,PV}, E_{c,BESS})$$

Under the same BESS constraints as defined before.

3.3.2 Intra-day offering strategy

Using our market structure model, the difference in revenue $R$ when an intra-day offer of volume $E_{ID}$ and price $\pi_{ID}$ is accepted writes:

$$R = \pi_s E - (E + E_{ID} - E_c)(\pi - \pi_B) + (\pi - \pi_{ID})E_{ID}$$

If $E_{ID} > 0$, the expected value of the revenue with respect to the PV power uncertainty is:

$$E(R) = \pi_s E(E) + (\pi_s - \pi_{ID})E_{ID}$$
$$-(\pi_s - \pi_-)\int_{-\infty}^{E_c - E_{ID}} (e + E_{ID} - E_c)f_{PV}(e)de$$
$$-(\pi_s - \pi_+)\int_{E_c - E_{ID}}^{+\infty} (e + E_{ID} - E_c)f_{PV}(e)de$$

By deriving with respect to $E_{ID}$ using the Leibniz integral rule, we get:

$$\frac{dE(R)}{dE_{ID}} = F_{PV}(E_c - E_{ID})(\pi_- - \pi_+) - (\pi_{ID} - \pi_+)$$

By equaling the derivative to zero, we find a critical point at:

$$E_{ID}^* = E_c - F_{PV}^{-1}\left(\frac{\pi_{ID} - \pi_+}{\pi_- - \pi_+}\right)$$

Note that the second derivative is:
$$-(\pi_- - \pi_+)f_{PV}(E_c - E_{ID})$$

This is always negative by definition of the balancing prices. Thus, this critical point is a local maximum of the revenue. Using the same method, we find a similar result for the case $E_{ID} > 0$.

The intra-day offering strategy for the maximization of revenue is thus to offer the optimal intra-day volume defined by this equation, using $\pi_{ID} = \pi_s$.

### 3.3.3 Real-time control

To perform the real-time control of the plant, we used the penalized revenue as the loss function of the MPC controller:

$$E_{BESS}^* = argmax_{\{E_{BESS} \in R^{N_{MPC}}\}} R'(E_{BESS})$$

Under the same BESS constraints as before.

## 4. RESULTS

We performed the control of the PV/BESS for imbalance minimization and revenue maximization over a period of 8 months from the first of September 2017 to the first of May 2018. The simulation has a temporal resolution of 30 minutes. We compared the results obtained for a single plant with 2.7 MWp and an aggregation of 13 plants with 98 MWp.

PV power forecasts and price forecasts were obtained respectively with an Analog Ensemble method (AnEn) and a Support Vector Machine (SVM) algorithm as in [12]. Weather data required for the forecasts was obtained from the European Center for Medium-range Weather Forecasts.

### 4.1 Imbalance minimization

For imbalance minimization, we assumed that we had a theoretical infinite storage capacity. We simulated the control with or without an intra-day market and with a stochastic or a deterministic method. Then, we checked a posteriori the actual size of the BESS that was requested to perform the control.

Figure 1 shows the imbalance reduction that we obtained for the single and aggregated plant. For each plant, we can distinguish three curves: one without intra-day market, which sets the benchmark for the BESS size, one with intra-day market and deterministic control, and one with intra-day market and stochastic control.

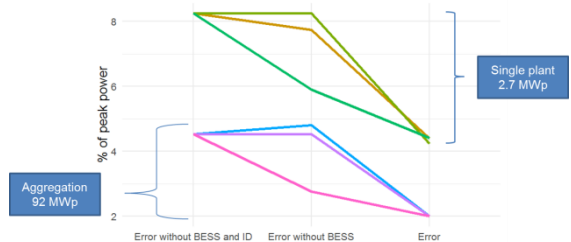

**Figure 1:** Imbalance reduction for the different control strategies

As expected, the aggregated plant has a lower initial imbalance compared to its installed power. All three strategies have the same resulting imbalance. Thus, when comparing the actual BESS size required for each strategy, we compare the actual efficiency in BESS usage of each strategy, since they all provide the same imbalance reduction.

Storage size requirements are reported on table I. We can see that the use of an intra-day market drastically reduces the required size of the BESS. However, there is not much difference between stochastic and deterministic control of the BESS.

**Table I:** BESS size requirements for imbalance minimization

|  | Single Plant | Aggregation |
|---|---|---|
| Required size for ideal control (MWh/MWp) | 45 | 28 |
| Actual size with ID and stochastic control (MWh/MWp) | 14.0 | 9.35 |
| Size reduction (%) | -68.9 | -66.7 |

### 4.2 Revenue Maximization

For revenue maximization, we cannot assume that we have a theoretical infinite capacity. In a case when discharging the BESS adds to the revenue, the optimization would not result in a finite solution, since the optimal output of the BESS would be infinite. Thus, we performed the control for a BESS size of 1 MWh/MWp and compared the revenue variation from all strategies.

Results are shown on table II. We can see that there is no strategy that increases the revenue. This is caused by the high uncertainty in balancing prices. In a significant portion of the simulation, imbalances that were helpful for the grid at a national level were remunerated. For example, if the producer overproduces while the power grid is short of energy, this imbalance is remunerated at a higher price than the spot price. Thus, compensating this forecast error is detrimental to the revenue. Since it is very difficult to forecast whether the grid will be in short or long in energy at the national level, using the BESS to compensate forecast errors is not reliable. In the end, there was no strategy that was able to increase the revenue.

Still, the methods using stochastic control were more efficient that the deterministic ones, as they included more information. However, they only used information on the uncertainty from the PV power production and not the spot and balancing prices. They were thus able to mitigate the high uncertainty of balancing prices but not enough to reliably increase the revenue.

**Table II:** Revenue increase for the different strategies

|  | Single plant | | Aggregation | |
|---|---|---|---|---|
| Revenue without BESS and ID (€) | 77490 | | 2323880 | |
|  | With ID | Without ID | With ID | Without ID |
| Price improvement – deterministic control (%) | -1.4 | -1.1 | -1.2 | -1.0 |
| Price improvement – stochastic control (%) | -0.77 | -0.48 | -0.4 | -1.0 |

## 5. CONCLUSIONS

In this paper, we studied the effect of intra-day markets and plant aggregation on the BESS size requirement of a PV/BESS system in an electricity market. We compared a deterministic and a stochastic control method for both imbalance reduction and revenue maximization.

We could show that the BESS size reduction obtained with plant aggregation is around 50 % when going from a single plant of 2.7 MWp to an aggregation of 13 plants that have an installed power of 98 MWp in total. Participating in intra-day electricity markets can also reduce the BESS size requirement up to 68 % in the best case, that is, when all intra-day offers are accepted. For imbalance reduction, the uncertainty in PV production is low enough on the short-term that there is little difference between stochastic and deterministic control methods.

On the other hand, for revenue maximization, stochastic control methods perform better, as they are able to somehow mitigate the high uncertainty of balancing prices. However, using the BESS always resulted in a lower income, because in a significant portion of the time, imbalances were actually remunerated by the TSO.